\documentclass[a4paper,12pt,amssymb,amsmath,tightenlines]{article}
\usepackage{amsmath, amssymb, amsthm, latexsym, mathrsfs}
\usepackage{dsfont}
\usepackage{color}

\newcommand{\rd}{\mbox{$\rm d$}}
\newcommand{\PR}{\mathbb{P}}
\newcommand{\Q}{\mathbb{Q}}

\newcommand{\Fcal}{\mathcal{F}}
\newcommand{\Gcal}{\mathcal{G}}

\newcommand{\E}{\mathbb{E}}
\newcommand{\B}{\mathbb{B}}
\newcommand{\EM}{\mathcal{E}}

\newcommand{\R}{\mathbb{R}}

\newcommand{\nn}{\nonumber}
\newtheorem{lem}{Lemma}[section]
\newtheorem{prop}{Proposition}[section]

\theoremstyle{definition}

\newtheorem{rem}{Remark}[section]
\numberwithin{equation}{section}

\title{\bf{Conditional Density Models \\ for Asset Pricing}}
\begin{document}

\author{Damir Filipovi\'c\,$^{\ast}$, Lane P. Hughston$^{\dag}$ \& Andrea Macrina\,$^{\ddag\,\S}$
}
\date{}
\maketitle
\begin{center}
$^{\ast}$ Swiss Finance Institute, Ecole Polytechnique F\'ed\'erale de Lausanne, Switzerland\\
$^{\dag}$ Department of Mathematics, Imperial College London, London SW7 2AZ, UK\\
$^{\ddag}$ Department of Mathematics, King's College London, London WC2R 2LS, UK\\
$^{\S}$ Institute of Economic Research, Kyoto University, Kyoto
606-8501, Japan
\end{center}
\begin{abstract}
We model the dynamics of asset prices and associated derivatives by
consideration of the dynamics of the conditional probability density
process for the value of an asset at some specified time in the
future. In the case where the price process is driven by Brownian motion, an
associated ``master equation" for the dynamics of the conditional
probability density is derived and expressed in integral form. By a
``model" for the conditional density process we mean a solution to
the master equation along with the specification of (a) the initial
density, and (b) the volatility structure of the density. The volatility structure is assumed at any time and for each value of the argument of the density to be a functional of the history of the density up to that time. In practice one specifies the
functional modulo sufficient parametric freedom to allow for the
input of additional option data apart from that implicit in the initial density. The scheme is sufficiently flexible to allow for the input of various types of data
depending on the nature of the options market
and the class of valuation problem being undertaken. Various
examples are studied in detail, with exact solutions
provided in some cases.
\begin{center}
9 November  2011
\end{center}
\end{abstract}
\begin{center}
Classification: C60, C63, G12, G13.
\\
\vspace{.25cm}
information-based asset pricing; nonlinear filtering; Breeden-Litzenberger
equation.
\\ \vspace{.25cm}
\end{center}
\section{Introduction}
This paper is concerned with modelling the dynamics of the volatility surface. The
problem is of great practical interest to traders, and as a consequence has an extensive mathematical literature associated with it. In this brief report, we shall not attempt to review earlier work in the area, but refer the reader, for example, to Sch\"onbucher (1999), Gatheral (2006), Schweizer \& Wissel (2008a,b), Carmona \& Nadtochiy (2009, 2011), and references therein. 
Put informally, the general idea of the paper is as follows. We fix a numeraire, and write $\{A_t\}$ for the value process of some tradable financial asset expressed in units of that numeraire. We fix a time $T$ and assume that no dividends are paid from time 0 up to $T$. Letting $\Q$ denote the martingale measure associated with the chosen numeraire, we have
\begin{equation}
A_t=\E^{\Q}\left[A_T\,\vert\,\Fcal_t\right],
\end{equation}
for $t\le T$, and 
\begin{equation}
 C_{tT}(K)=\E^{\Q}\left[\left(A_T-K\right)^+\,\vert\,\Fcal_t\right],
\end{equation}
where $C_{tT}(K)$ denotes the price at time $t$ of a $T$-maturity call option with strike $K$. The associated conditional density process $\{f_t(x)\}$ for the random variable $A_T$ is defined by
\begin{equation}\label{defcondtionaldensity}
\int^y_{-\infty}f_t(x)\rd x=\E^{\Q}\left[{\mathds 1}\{y>A_T\}\,\vert\,\Fcal_t\right]. 
\end{equation}
Then for the asset price we have 
\begin{equation}\label{assetpriceproc}
 A_t=\int_{\R}x f_t(x)\rd x,
\end{equation}
and the corresponding option prices are given by
\begin{equation}\label{optionproc}
 C_{tT}(K)=\int_{\R}(x-K)^+f_t(x)\rd x.
\end{equation}
Instead of modelling $\{A_t\}$  and then determining $\{C_{tT}(K)\}$, our strategy is to model the conditional density process. Then the underlying asset price process and the associated option prices are determined by (1.4) and (1.5). Roughly speaking, the idea is to model $\{f_t(x)\}$ in such a way that it contains some parametric freedom that can be calibrated to a specified range of initial option prices. Models for conditional densities have been considered in various contexts in finance. These include for example applications to interest rates (Brody \& Hughston 2001a,b, 2002, Filipovi\'c {\it et al.}~2010), and to credit risk (El Karoui {\it et al}.~2010). Although mostly different from what has previously appeared in the literature, our approach to modelling the volatility surface is similar in spirit in some respects to that of Davis (2004). 

Let us consider in more detail the class of assets that will form the basis of our investigation. We introduce a probability space $(\Omega,\Fcal,\PR)$ with filtration $\{\Fcal_t\}_{t\ge 0}$, where $\PR$ is the ``physical'' measure and $\{\Fcal_t\}$ is the market filtration. We assume that price processes are adapted to $\{\Fcal_t\}$. We assume the absence of arbitrage, and the existence of an established pricing kernel $\{\pi_t\}$ associated with some choice of base currency as numeraire. We work in the setting of a multi-asset market, and do not assume that the market is complete.

We write $\{A^i_t\}_{i=0,1,\ldots,N}$ for the price
processes of a collection of non-dividend-paying tradable financial assets. Prices are
expressed in units of the base currency. We
refer to asset $i$ as $A^i$. We model the $\{A_t^i\}$ as Ito
processes, and for each $A^i$ we require that 
\begin{equation}
\pi_s A^i_s=\E^{\PR}\left[\pi_t A^i_t\,\vert\,\Fcal_s\right]
\end{equation}
for $s\le t$. Such an asset is characterized by its value $A^i_T$ at some terminal date $T$. In some situations it is useful to regard the asset as offering a single payment at $T$.  In that case $\{A^i_t\}_{0\le t<T}$ represents the price process of the asset that offers such a payment. In other situations we can consider $A_T$ as being a ``snapshot'' of the value of the asset at time $T$. Usually the context will make it clear which meaning is intended.

Let $\{A^0_t\}$ be a money-market account in the base currency, initialised to
unity. If we set $\rho_t=\pi_t A^0_t$ for $t\ge 0$, it follows that
$\{\rho_t\}$ is a $\PR$-martingale. A standard argument shows that
$\{\rho_t\}$ can be used to make a change of measure. The resulting
measure $\Q^0$ is the ``risk-neutral" measure associated with the
base currency, and has the property that if the price of any
non-dividend-paying asset is expressed in units of the money-market
account, then the resulting process is a $\Q^0$-martingale. Thus, for each $i$ we have
\begin{equation}
A^i_s=A^0_s\,\E^{\Q^0}\left[\frac{A^i_t}{A^0_t}\,\bigg\vert\,\Fcal_s\right].
\end{equation}

A similar situation arises with other choices of numeraire.
Specifically, for any non-dividend-paying asset $A^i$ of
limited liability ($A^i_t>0$), with price
$\{A^i_t\}$, there is an associated measure $\Q^i$ with the property
that if the price of any non-dividend-paying asset is expressed in units of $A^i$ then
the result is a $\Q^i$-martingale. Thus for $0\le s\le t$ and for all $i,j$ for which the price of $A^i$ is strictly positive we have:
\begin{equation}
A^j_s=A^i_s\E^{\Q^i}\left[\frac{A^j_t}{A^i_t}\,\bigg\vert\,\Fcal_s\right].
\end{equation}

Bearing these points in mind, we observe that the
option pricing problem can be formulated in the following
context. We consider European-style options of a
``Margrabe'' type, for a pair of non-dividend-paying assets $A^i$
and $A^j$, where the option-holder has the right at time $t$ to
exchange $K$ units of $A^i$ for one unit of $A^j$. The payoff $H_t^{ij}$ of such an option, in units of the base currency, is of the form
\begin{equation}
H^{ij}_t(K)=\left(A^j_t-KA^i_t\right)^+.
\end{equation}
The value of the option at $s\le t$, expressed
in units of the base currency, is given by
\begin{equation}
C_{st}^{ij}(K)=\frac{1}{\pi_s}\E^{\PR}\left[\pi_t\left(A^j_t-KA^i_t\right)^+\,\vert\,\Fcal_s\right].
\end{equation}
If $A^i$ is of limited liability, then the option value, expressed in units of $A^i$, is a $\Q^i$-martingale:
\begin{equation}
\frac{C_{st}^{ij}(K)}{A^i_s}=\E^{\Q^i}\left[\left(\frac{A^j_t}{A^i_t}-K\right)^+\,\bigg\vert\,\Fcal_s\right].
\end{equation}

This relation can be expressed more compactly as follows. Write $A_s$ for the price at time $s\le t$ of a ``generic asset" expressed in units of a ``generic numeraire", and $C_{st}(K)$ for the
price, in units of the chosen numeraire, at time $s\le t$,
of a $t$-maturity $K$-strike option. Then the option payoff is given (in numeraire units) by
\begin{equation}
H_t(K)=\left(A_t-K\right)^+
\end{equation}
The value of the option at time $s\le t$ is
\begin{equation}\label{svalueoption}
C_{st}(K)=\E^{\Q}\left[\left(A_t-K\right)^+\,\vert\,\Fcal_s\right],
\end{equation}
where $\Q$ is the martingale measure associated with the
numeraire. By ``generic'' we mean any choice of a non-dividend-paying assets $A^i$ and $A^j$ such that $A^i$ is of limited liability. 

Standard options are not included in the category
discussed above. A standard call has the payoff
$H_t=\left(A_t-K\right)^+$ where $A_t$ is the price of the
underlying at time $t$ in currency units, and $K$ is a fixed
strike in currency units. This is not an option
to exchange $K$ units of a non-dividend-paying asset for one unit of
another non-dividend-paying asset. One should think of the
fixed strike as being $K$ units of a unit floating rate note.
The only asset that maintains a constant value (in
base-currency units) is a floating rate note---and such an asset
pays a dividend. The dividend is the interest rate. In a given
currency, by a floating rate note we mean an idealised note that
pays interest continuously (rather than in lumps). The associated dividend is the short
rate. One might argue that the strike on a
standard option is cash, and that cash is a non-dividend-paying
asset: this point of view leads to paradoxes. In the standard
theory we regard cash as paying an implicit dividend,
a convenience yield, in the form of a liquidity benefit
equivalent to the interest rate. In summary, a standard option is an option to exchange the asset with a floating-rate note, which pays a ``dividend''.

Thus a standard option is a complicated entity---it is an option to exchange a certain number of
units of a dividend-paying asset for one unit of a
non-dividend-paying asset. It is more logical first to
examine an option based on a pair of {\it
non}-dividend-paying assets. In the literature this approach is
implicitly adopted through the device of ``setting the
interest rate equal to zero''. In that situation the floating rate
note is non-dividend-paying; thus, the setting we
operate within includes the zero-interest case. One would like to tackle the general problem of an option to
exchange $K$ units of one dividend-paying asset for one unit of
another dividend-paying asset (a ``standard" foreign exchange option
falls into that category); but, unless the dividend (or the
interest rate) systems are deterministic, this is a more difficult
problem than the one we propose to consider here.

The structure of the paper is as follows. In Section 2 we derive a dynamical equation  for the conditional density, which we call the ``master equation'', given by (\ref{ME}). An integral form of the equation is presented in (\ref{f-intform}), which forms the basis of the solutions presented in later sections. In Section 3 we specify the general form we require the volatility structure of the conditional density to take, and give a characterization of what we mean by a ``conditional density model'' for asset pricing. In Section 4 we consider in detail the class of models for which the volatility structure of the conditional density is a deterministic function of two variables. This family of models admits a complete solution by use of a filtering technique. The resulting asset prices exhibit a stochastic volatility that is adapted to the market filtration but that is not in general of the local-volatility type. In Section 5 we consider the case when the volatility structure is linear in the terminal value of the asset. In that case the resulting models are Markovian, and can be calibrated to an arbitrary initial density. In Section 6 an alternative representation of the semi-linear case is presented using a Brownian-bridge technique. In Section 7 we show that the Bachelier model and the geometric Brownian motion model arise as special cases of the semilinear models, for particular choices of the initial density.  We conclude in Section 8 with the calculation of option prices.

\section{Conditional density processes}
The market is understood as having the setup
described in the previous section. We have a probability space
$(\Omega,\Fcal,\Q)$ with filtration $\{\Fcal_t\}$. A non-dividend-paying limited-liability asset is chosen as numeraire, and all prices are expressed in units of that numeraire. The
measure $\Q$ has the property that the price process of a
non-dividend-paying asset, when expressed in units of the
numeraire, is a martingale. We refer to $\Q$ as the martingale
measure associated with this numeraire.

We fix  $T>0$, and assume for $0\le t< T$ the existence of an
$\Fcal_t$-conditional $\Q$-density $f_{t}(x)$ for $A_T$. In our applications we
have in mind the cases $x\in\R$ and $x\in\R^+$,
but it is useful to be flexible as regards the
choice of the domain of the density function. In what follows we treat the case $x\in\R$, and leave it to the reader to supply the necessary adjustments for other domains. Thus we assume the existence of a density process $\{f_t(x)\}$, $x\in\R$, $0\le t<T$, such that (\ref{defcondtionaldensity}) holds. It follows that for any bounded, measurable function $g(x)$, $x\in\R$, we have  
\begin{equation}
\int_{\R} g(x)f_{t}(x)\rd
x=\E\left[g(A_T)\,\vert\,\Fcal_t\right].
\end{equation}
We ask that $\{f_t(x)\}$ should have those properties that follow heuristically as consequence of the formula obtained by formally differentiating (\ref{defcondtionaldensity}), namely: 
\begin{equation}
 f_t(x)=\E\left[\delta(x-A_T)\,\vert\,\Fcal_t\right].
\end{equation}
In particular, we require the following: (a) that for each $x\in\R$ the process $\{f_t(x)\}_{0\le t<T}$ is an $\{\Fcal_t\}$-martingale, and hence $f_{s}(x)=\E\left[f_{t}(x)\,\vert\,\Fcal_s\right]$ for $0\le s\le t<T$; (b) that  the price of the asset can be expressed in terms of the density by (\ref{assetpriceproc})
for $t<T$; and (c) that
\begin{equation}
 \lim_{t\rightarrow T}\int_{\R}x\,f_{t}(x)\rd x=A_T.
\end{equation}
We also assume, where required, that expressions analogous to (\ref{assetpriceproc}) can be written for claims
based on $A_T$. For example, if $C_{tT}(K)$ denotes the price at $t$
of a $T$-maturity, $K$-strike European call option, then we assume that (\ref{optionproc}) holds.

In the applications that follow, we introduce a $\Q$-Brownian motion $\{W_t\}$, which we take to be adapted to $\{\Fcal_t\}$, and specialize to the case for which the dynamical equation of $\{f_t(x)\}$ is of the form 
\begin{equation}\label{Intdynamics}
f_t(x)=f_0(x)+\int^t_0\sigma^f_s(x)\,f_s(x)\rd W_s, 
\end{equation}
for some process $\{\sigma^f_t(x)\}$, $x\in\R$, $0\le t<T$, representing the volatility of the density. It follows that
\begin{equation}
 A_t=A_0+\int^t_0\sigma_s^A \rd W_s,
\end{equation}
where 
$A_0=\int_{\R}xf_0(x)\rd x$,
and where the volatility of $A_t$ is given by
\begin{equation}
 \sigma^A_t=\int_{\R}x\sigma^f_t(x)f_t(x)\rd x.
\end{equation}
For simplicity, we consider in this paper the case where $\{W_t\}$ is a one-dimensional Brownian motion. The extension to the multi-dimensional situation is unproblematic.
\begin{lem} 
\rm{
 The normalisation condition 
\begin{equation}\label{normalization}
\int_{\R}f_t(x)\rd x=1
\end{equation}
holds for all $t\in[0,T)$ if and only if there exists a process $\{\sigma_t(x)\}$ such that 
\begin{equation}\label{volmodel}
\sigma^f_t(x)=\sigma_t(x)\int_{\R}f_t(y)\rd y-\int_{\R}\sigma_t(y)f_t(y)\rd y
\end{equation}
for almost all $t\in[0,T)$, and we have the initial condition
\begin{equation}\label{initialnorm}
 \int_{\R}f_0(x)\rd x=1.
\end{equation}
}
\end{lem}
\noindent{\bf Proof}. First we show that (\ref{volmodel}) and (\ref{initialnorm}) imply (\ref{normalization}). Starting with (\ref{Intdynamics}), we integrate with respect to $x$ to obtain
\begin{equation}
 \int_{\R}f_t(x)\rd x=\int_{\R}f_0(x)\rd x+\int^t_0\int_{\R}\sigma^f_s(x)f_s(x)\rd x\,\rd W_s.
\end{equation}
Inserting (\ref{volmodel}) and using (\ref{initialnorm}) we obtain (\ref{normalization}). Conversely, if we assume (\ref{normalization}) then (\ref{initialnorm}) holds as well, and hence
\begin{equation}
 \int^t_0\int_{\R}\sigma^f_s(x)f_s(x)\rd x\,\rd W_s=0,
\end{equation}
from which it follows that 
\begin{equation}
 \int_{\R}\sigma^f_t(x)f_t(x)\rd x=0
\end{equation}
for almost all $t\in[0,T)$, and thus that  (\ref{volmodel}) holds for almost all  $t\in[0,T)$ for some $\sigma_t(x)$.
\hfill$\Box$
\\

Thus, once we specify $\{\sigma_{t}(x)\}$ and
$\{f_{0}(x)\}$, the dynamical equation for the density---the so-called master
equation---takes the form
\begin{equation}\label{ME}
f_{t}(x)=f_{0}(x)+\int^t_0\left[\sigma_{s}(x)-\int_{\R}\sigma_{s}(y)f_{s}(y)\rd
y\right]f_s(x)\,\rd W_s.
\end{equation}
\begin{lem}\label{lemmastereq} 
\rm{
The conditional desity process $\{f_t(x)\}$ satisfies the master equation (\ref{ME}) with initial density $\{f_{0}(x)\}$ and volatility structure $\{\sigma_{t}(x)\}$ if and only if
\begin{equation}\label{f-intform}
f_{t}(x)=\frac{f_0(x)\exp\left(\int^t_0\sigma_s(x)\rd
Z_s-\tfrac{1}{2}\int^t_0{\sigma_s}^2(x)\rd
s\right)}{\int_{\R}f_0(y)\exp\left(\int^t_0\sigma_{s}(y)\rd
Z_s-\tfrac{1}{2}\int^t_0{\sigma_s}^2(y)\rd s\right)\rd y},
\end{equation}
where
\begin{equation}\label{ProcessZ}
Z_t=W_t+\int^t_0\int_{\R}\sigma_s(y)f_s(y)\rd y\,\rd s.
\end{equation}
}
\end{lem}
\noindent{\bf Proof}. Writing (\ref{ME}) in differential form, we have
\begin{equation}\label{f-diffform}
\rd f_t(x)=f_t(x)\left(\sigma_t(x)-\langle\sigma_t\rangle\right)\rd W_t\,,
\end{equation}
where for convenience we write
\begin{equation}\label{bracketnotation}
\langle\sigma_{t}\rangle=\int_{\R}\sigma_{t}(x)f_{t}(x)\rd x
\end{equation}
for the conditional ``mean'' of the volatility. We integrate (\ref{f-diffform}) to obtain
\begin{align}
f_{t}(x)=f_{0}(x)\exp\left[\int^t_0\left(\sigma_{s}(x)-\langle\sigma_{s}\rangle\right)\rd
W_s-\tfrac{1}{2}\int^t_0\left(\sigma_{s}(x)-\langle\sigma_{s}\rangle\right)^2\rd
s\right].
\end{align}
Expanding the exponent, we have
\begin{eqnarray}
f_{t}(x)=f_{0}(x)\frac{\exp\left[\int^t_0\sigma_{s}(x)\left(\rd
W_s+\langle\sigma_{s}\rangle\rd
s\right)-\tfrac{1}{2}\int^t_0{\sigma_{s}}^2(x)\rd
s\right]}{\exp\left[\int^t_0\langle\sigma_{s}\rangle\left(\rd
W_s+\langle\sigma_{s}\rangle\rd
s\right)-\tfrac{1}{2}\int^t_0\langle\sigma_{s}\rangle^2\rd
s\right]}.
\end{eqnarray}
Then we introduce a process $\{Z_t\}$ by writing
\begin{equation}\label{Z-process}
Z_t=W_t+\int^t_0\langle\sigma_{s}\rangle\rd s,
\end{equation}
and it follows that
\begin{eqnarray}\label{Z-density}
f_{t}(x)=f_{0}(x)\frac{\exp\left[\int^t_0\sigma_{s}(x)\rd
Z_s-\tfrac{1}{2}\int^t_0{\sigma_{s}}^2(x)\rd
s\right]}{\exp\left[\int^t_0\langle\sigma_{s}\rangle \rd
Z_s-\tfrac{1}{2}\int^t_0\langle\sigma_{s}\rangle^2\rd s\right]}.
\end{eqnarray}
Applying the normalization condition
(\ref{normalization}) to equation (\ref{Z-density}) above, we see that
\begin{align}\label{Trick}
&\exp\left(\int^t_0\langle\sigma_{s}\rangle \rd
Z_s-\tfrac{1}{2}\int^t_0\langle\sigma_{s}\rangle^2\rd
s\right)\nn\\
&\hspace{2cm}=\int_{\R}f_{0}(x)\exp\left(\int^t_0\sigma_{s}(x)\rd
Z_s-\tfrac{1}{2}\int^t_0{\sigma_{s}}^2(x)\rd s\right)\rd x.
\end{align}
It follows that (\ref{Z-density}) reduces to
(\ref{f-intform}). Conversely, if $f_t(x)$ is given by (\ref{f-intform}), then it is a straightforward exercise in Ito calculus to check that the master equation is satisfied.
\hfill$\Box$
\section{Conditional density models}
We are in a position now to say more precisely what we mean by a ``conditional
density model''. In doing so, we are motivated in part by
advances in the study of infinite-dimensional stochastic
differential equations. By a ``model'' for the density process we
understand the following. We consider solutions of the master
equation (\ref{ME}) satisfying the normalization condition
(\ref{normalization}), in conjunction with the specification of: (a) the {\it initial density} $f_{0}(x)$; and (b) the {\it volatility structure} $\{\sigma_{t}(x)\}$ in the form of a
functional
\begin{equation}
\sigma_{t}(x)=\Phi[f_{t}(\cdot), t, x].
\end{equation}
For each $x$ and $t$ the volatility $\sigma_{t}(x)$ depends on
$f_{t}(y)$ for all $y\in\R$. Hence (\ref{ME}), thus specified,
determines the dynamics of an infinite-dimensional Markov process.

The initial density $f_{0}(x)$ can be determined if one supplies initial
option price data for the maturity date $T$ and for all strikes
$K\in\R$. In particular, we have
\begin{eqnarray}
C_{0T}(K) = \E[(A_T-K)^+]
= \int_{\R}(x-K)^+f_{0}(x)\rd x.
\end{eqnarray}
By use of the idea of Breeden \& Litzenberger (1978) we see, in the
present context, that for each value of $x$ one has
\begin{equation}
f_{0}(x)=\frac{\partial^2 C_{0T}(x)}{\partial x^2}.
\end{equation}
Here $f_{0}(x)$ is not generally the risk-neutral density, but rather the
$\Q$-density of the value at time $T$ of the asset in units of
the chosen numeraire. For simple practical applications, we take---in common with much of the literature---the numeraire to the be the money-market account, and set the interest rate to zero. Then $f_0(x)$ is the risk-neutral density.  Once $f_{0}(x)$ has been supplied, the
choice of the functional $\Phi$ determines the model
for the conditional density: we give some examples later in the
paper.

 In practice, one would like to specify $\Phi$ modulo
sufficient parametric freedom to allow the input of additional
option price data. What form this additional data might take depends
on the nature of the market and the class of
valuation problems being pursued. For example, a standard problem would be to look at a
limited-liability asset and consider additional data in the form of
initial option prices for all strikes in $\R^+$ and all maturities
in the strip $0< t\le T$. We require that $\Phi$ should be
specified in such a way that once the data are provided, then $\Phi$
is determined and the ``master equation'' provides an
evolution of the conditional density. Once we have the conditional
density process, we can work out the evolution of the option price
system for the specified strip, and hence the evolution of the
associated implied volatility surface. 

The data do not have to be presented exactly in the way 
specified in the previous paragraph---there may be
situations where more data are available (e.g., in the form of
barrier option prices or other derivative prices) or where less data
are available (less well-developed markets). One should think of the
parametric form of $\Phi$ as being adapted in a flexible way to the
nature of a specific problem. The philosophy is that there are many
different markets for options, and one needs a methodology that can
accommodate these with reasonable generality. 

It should be evident that an arbitrary solution to the master equation need not be a density process for an $\Fcal_t$-measurable random variable---additional assumptions are required concerning the nature of the volatility structure in order to ensure that $f_t(x)$ converges in an appropriate sense to a suitable Dirac distribution. In the examples given, we indicate how this can be achieved, in various situations, by the choice of the volatility structure.
\section{Models with deterministic volatility structures}\label{Sec Imp Dens Mod}
We proceed to present a rather general class of conditional density models characterized by a deterministic volatility structure. These are models for which $\{\sigma_t(x)\}$ is of the form 
\begin{equation}
\sigma_t(x)=v(t,x) 
\end{equation}
for some deterministic function $v(t,x)$ defined for appropriate values of $t$ and $x$. We are able to give a more or less complete construction of such models in the form of a ``weak'' solution of the master equation. By a weak solution we mean that on a probability space $(\Omega,\Fcal,\Q)$ we construct a filtration $\{\Fcal_t\}$, a Brownian motion $\{W_t\}$, and a conditional density process $\{f_t(x)\}$ satisfying the master equation and having the desired properties.
More specifically, let $T\in(0,\infty)$ be fixed, and let $f_0(x):\R\rightarrow\R^+$ be a prescribed initial density. Let $v(t,x)$ be a function on $[0,T)\times\R$ satisfying
\begin{equation}
\int^t_0v(s,x)^2 \,\rd s < \infty
\end{equation}
for $t<T$ and $x \in \R$, and let $\gamma(t)$ be a function on $[0,T)$ such that 
\begin{equation}\label{measurecond}
 \lim_{t\rightarrow T}\gamma(t)=0, \quad  \quad  \lim_{t \rightarrow T}\gamma(t)\int^t_0v(s,x)\rd s=g(x),
\end{equation}
for some invertible function $g(x)$ on $\R$. We have the following:
\begin{prop}\label{Prop inftime models}{\rm Let $X$ have density $f_0(x)$, and let $\{B_t\}$ be an independent Brownian motion. Let $\{\Fcal_t\}$
be the filtration generated by the ``information'' process $\{I_t\}$ defined by
\begin{equation}\label{IDef}
I_t=B_t+\int^t_0 v(s,X)\rd s.
\end{equation}
Let $\{f_t(x)\}$ be defined by
\begin{equation}\label{PROPFCOND}
f_{t}(x)=\frac{f_{0}(x)\exp\left[\int^t_0 v(s,x)\rd
I_s-\tfrac{1}{2}\int^t_0 v^2(s,x)\rd
s\right]}{\int^{\infty}_{-\infty}f_{0}(y)\exp\left[\int^t_0
v(s,y)\rd I_s-\tfrac{1}{2}\int^t_0 v^2(s,y)\rd s\right]\rd y},
\end{equation}
and define $\{W_t\}$ by setting
\begin{equation}\label{XInnovations}
W_t=I_t-\int^t_{0}\E^{\Q}\left[v(s,X)\,\vert\,\Fcal_s\right]\rd s.
\end{equation}
Then: (a) the random variable $X$ is $\Fcal_T$-measurable, and $f_t(x)$ is the associated conditional density; (b) the process
$\{W_t\}$ is an $\{\Fcal_t\}$-adapted Brownian motion; and (c) for $t\in[0,T)$, the density process $\{f_t(x)\}$ satisfies the master equation
\begin{equation}
f_t(x)=f_0(x)+\int^t_0
f_s(x)\left[v(s,x)-\int^{\infty}_{-\infty}v(s,y)f_s(y)\rd
y\right]\rd W_s\,.\label{fWequation}
\end{equation}
}
\end{prop}
\noindent{\bf Proof}. It follows from (\ref{measurecond}) that $g(X)=\lim_{t \rightarrow T}\gamma(t) I_t$ is $\Fcal_T$-measurable. Since $g(x)$ is invertible, we conclude that $X$ is $\Fcal_T$-measurable. Now let the filtration $\{\Gcal_t\}$ be defined by
\begin{equation}
\Gcal_t=\sigma\left(\{B_s\}_{0\le s\le t}, X\right).
\end{equation}
Clearly $\Fcal_t\subset\Gcal_t$. The random variable $X$ is
$\Gcal_t$-measurable, and $\{B_t\}$ is a $(\{\Gcal_t\},\Q)$-Brownian
motion. We introduce a $(\{\Gcal_t\},\Q)$-martingale $\{M_t\}$ 
by setting
\begin{equation}
M_t=\exp\left[-\int^t_0 v(s,X_{})\rd B_s-\tfrac{1}{2}\int^t_0
v^2(s,X_{})\rd s\right],
\end{equation}
for $t\in [0,T)$, and we let the probability measure $\B$ be defined by
\begin{equation}\label{B measure}
\frac{\rd\B}{\rd\Q}\bigg\vert_{\Gcal_t}=M_t.
\end{equation}
We observe that $\rd I_t=\rd B_t+v(t,X_{})\rd t$. By Girsanov's theorem, $\{I_t\}$ is a
$(\{\Gcal_t\},\B)$-Brownian motion. We note that $\{M^{-1}_t\}$
is a $(\{\Gcal_t\},\B)$-martingale. Let $H$ be a bounded measurable function on $\R$. Since $H(X)$ is $\Gcal_t$-measurable, we
have the generalised Bayes formula
\begin{equation}\label{Bcondexpect}
\E^{\Q}\left[H(X_{})\vert\,\Fcal_t\right]=\frac{\E^{\B}\left[M^{-1}_t
H(X_{})\,\vert\,\Fcal_t\right]}{\E^{\B}\left[M^{-1}_t\,\vert\,\Fcal_t\right]}.
\end{equation}
Let us work out the right-hand side of this
equation. To this end we show that $X_{}$ and $I_t$ are
$\B$-independent for all $t$. In particular, we show that the
generating function
\begin{equation}\label{Charactfct}
\E^{\B}\left[\exp\left(yI_t+zX_{}\right)\right]
\end{equation}
factorises. We have:
\begin{align}
\E^{\B}\left[\exp\left(yI_t+zX_{}\right)\right]&=\E^{\Q}\left[M_t\exp\left(yI_t+zX_{}\right)\right], \\
&=\E^{\Q}\left[\exp\left(-\int^t_0 v(s,X_{})\rd
B_s-\tfrac{1}{2}\int^t_0 v^2(s,X_{})\rd s\right.\right.\nn \\
&\left.\left.\hspace{2.5cm}+y\left[B_t+\int^t_0 v(s,X_{})\rd
s\right]\right)\exp\left(zX_{}\right)\right],\label{intromart}  \\
&=\E^{\Q}\left[m_t\,\exp\left(\tfrac{1}{2}y^2
t\right)\exp\left(zX_{}\right)\right],\label{expresmart}
\end{align}
where
\begin{equation}
m_t=\exp\left(\int^t_0\left[-v(s,X_{})+y\right]\rd
B_s-\tfrac{1}{2}\int^t_0\left[-v(s,X_{})+y\right]^2\rd s\right).
\end{equation}
By use of the tower property and the independence of $\{B_t\}$ and
$X_{}$ under $\Q$, we have
\begin{eqnarray}
\E^{\B}\left[\exp\left(yI_t+zX_{}\right)\right]&=&\E^{\Q}\left[m_t\exp\left(\tfrac{1}{2}y^2
t\right)\exp\left(z X_{}\right)\right],\\
&=&\exp\left(\tfrac{1}{2}y^2
t\right)\E^{\Q}\left[\E^{\Q}\left[m_t\exp\left(z
X_{}\right)\,\vert\,X_{}\right]\right],\\
&=&\exp\left(\tfrac{1}{2}y^2
t\right)\E^{\Q}\left[\E^{\Q}\left[m_t\,\vert\,X_{}\right]\E^{\Q}\left[\exp\left(zX_{}\right)\right]\right].
\end{eqnarray}
One observes that $\E^{\Q}[m_t\,\vert\,X_{}]=1$.
Thus we obtain the desired factorization:
\begin{equation}
\E^{\B}\left[\exp\left(yI_t+zX_{}\right)\right]=\exp\left(\tfrac{1}{2}y^2
t\right)\E^{\Q}\left[\exp\left(zX_{}\right)\right].
\end{equation}
Now that we have shown that $X_{}$ is $\B$-independent of $I_t$ (and
thus of $\Fcal_t$), we can work out the right-hand side
of (\ref{Bcondexpect}). We have:
\begin{eqnarray}
\E^{\Q}\left[H(X_{})\,\vert\,\Fcal_t\right]&=&\frac{\E^{\B}\left[M^{-1}_t
H(X_{})\,\vert\,\Fcal_t\right]}{\E^{\B}\left[M^{-1}_t\,\vert\,\Fcal_t\right]},\\
&=&\frac{\E^{\B}\left[H(X_{})\exp\left(\int^t_0 v(s,X_{})\rd
I_s-\tfrac{1}{2}\int^t_0 v^2(s,X_{})\rd
s\right)\,\vert\,\Fcal_t\right]}{\E^{\B}\left[\exp\left(\int^t_0
v(s,X_{})\rd I_s-\tfrac{1}{2}\int^t_0 v^2(s,X_{})\rd
s\right)\,\vert\,\Fcal_t\right]},\\
&=&\frac{\int^{\infty}_{-\infty}f_0(x)H(x)\exp\left(\int^t_0
v(s,x)\rd I_s-\tfrac{1}{2}\int^t_0 v^2(s,x)\rd s\right)\rd
x}{\int^{\infty}_{-\infty}f_0(y)\exp\left(\int^t_0 v(s,y)\rd
I_s-\tfrac{1}{2}\int^t_0 v^2(s,y)\rd s\right)\rd y}.
\end{eqnarray}
In particular, setting $H(X_{})={\bf 1}(X_{}\le x)$, we deduce
that $f_t(x)$ is the $\Fcal_t$-conditional density of $X$, as required. That proves the first part of the proposition. 

Next we show that $\{W_t\}$ is an $(\{\Fcal_t\},\Q)$-Brownian motion. We need to show (1) that $(\rd
W_t)^2=\rd t$, and (2) that $\E^{\Q}[W_u\,\vert\,\Fcal_t]=W_t$
for $0\le t\le u$. The first condition is evidently satisfied. The
second condition can be shown to be satisfied as follows. For
simplicity, we suppress the superscript $\Q$. We have
\begin{equation}\label{proof}
\E[W_u\,\vert\,\Fcal_t]=\E\left[I_u\,\vert\,\Fcal_t\right]-\E\left[\int^u_0
\E\left[v(s,X_{})\,\vert\,\Fcal_s\right]\rd
s\,\big\vert\,\Fcal_t\right].
\end{equation}
First we work out $\E\left[I_u\,\vert\,\Fcal_t\right]$. Since $\{B_t\}$ is a
$(\{\Gcal_t\},\Q)$-Brownian motion, we have
\begin{align}
\E\left[I_u\,\vert\,\Fcal_t\right]&=\E\left[B_u+\int^u_0
v(s,X_{})\rd
s\,\big\vert\,\Fcal_t\right]\\
&=\E\left[B_u\,\vert\,\Fcal_t\right]+\E\left[\int^u_0 v(s,X_{})\rd
s\,\big\vert\,\Fcal_t\right]\\
&=\E\left[\E\left[B_u\,\big\vert\,\Gcal_t\right]\,\big\vert\,\Fcal_t\right]+\E\left[\int^u_0
v(s,X_{})\rd
s\,\big\vert\,\Fcal_t\right]\\
&=\E\left[B_t\,\big\vert\,\Fcal_t\right]+\E\left[\int^u_0
v(s,X_{})\rd s\,\big\vert\,\Fcal_t\right]\label{I}.
\end{align}
We insert this intermediate result in (\ref{proof}) to obtain
\begin{align}
\E[W_u\,\vert\,\Fcal_t]&=\E\left[I_u\,\vert\,\Fcal_t\right]-\E\left[\int^u_0
\E\left[v(s,X_{})\,\vert\,\Fcal_s\right]\rd s\,\big\vert\,\Fcal_t\right],\\
&=\E\left[B_t\,\big\vert\,\Fcal_t\right]+\E\left[\int^u_0
v(s,X_{})\rd s\,\big\vert\,\Fcal_t\right]-\E\left[\int^u_0
\E\left[v(s,X_{})\,\vert\,\Fcal_s\right]\rd
s\,\big\vert\,\Fcal_t\right].
\end{align}
Next we split the integrals in the last two expectations by writing,
\begin{align}
\E[W_u\,\vert\,\Fcal_t]&=\E\left[B_t\,\big\vert\,\Fcal_t\right]+\E\left[\int^t_0
v(s,X_{})\rd s\,\big\vert\,\Fcal_t\right]+\E\left[\int^u_t
v(s,X_{})\rd
s\,\big\vert\,\Fcal_t\right]\nn\\
&-\E\left[\int^t_0 \E\left[v(s,X_{})\,\vert\,\Fcal_s\right]\rd
s\,\big\vert\,\Fcal_t\right]-\E\left[\int^u_t
\E\left[v(s,X_{})\,\vert\,\Fcal_s\right]\rd
s\,\big\vert\,\Fcal_t\right].
\end{align}
Observing that
\begin{equation}
\E\left[B_t\,\big\vert\,\Fcal_t\right]+\E\left[\int^t_0 v(s,X_{})\rd
s\,\big\vert\,\Fcal_t\right]=I_t,
\end{equation}
and that
\begin{equation}
\E\left[\int^u_t v(s,X_{})\rd
s\,\big\vert\,\Fcal_t\right]=\E\left[\int^u_t
\E\left[v(s,X_{})\,\vert\,\Fcal_s\right]\rd
s\,\big\vert\,\Fcal_t\right],
\end{equation}
we see that the expectation $\E[W_u\vert\,\Fcal_t]$ reduces to
\begin{align}
\E[W_u\,\vert\,\Fcal_t] =I_t-\E\left[\int^t_0
\E\left[v(s,X_{})\,\vert\,\Fcal_s\right]\rd
s\,\big\vert\,\Fcal_t\right]
 =I_t-\int^t_0 \E\left[v(s,X_{})\,\vert\,\Fcal_s\right]\rd s
 =W_t \,.
\end{align}
That shows that $\{W_t\}$ is an $\{\Fcal_t\}$-Brownian motion. An application of Ito calculus shows that the density process (\ref{PROPFCOND}) satisfies the SDE (\ref{fWequation}).\hfill$\Box$
\begin{rem}
For simplicity we have presented Proposition \ref{Prop inftime models} for the case of a one-dimensional state space. It should be clear from the proof how the results carry over to conditional densities on higher-dimensional state spaces.
\end{rem}
\begin{rem}
For simulations of the dynamics of the conditional
density process, the following alternative representation for (\ref{PROPFCOND}) may prove
useful:
\begin{equation}\label{altCondDens}
f_t(x)=\frac{f_0(x)\exp\left(\int^t_0\left[v(s,x)-v(s,X_{})\right]\rd
B_s-\tfrac{1}{2}\int^t_0\left[v(s,x)-v(s,X_{})\right]^2\rd s\right)}
{\int_{\R}f_0(y)\exp\left(\int^t_0\left[v(s,y)-v(s,X_{})\right]\rd
B_s-\tfrac{1}{2}\int^t_0\left[v(s,y)-v(s,X_{})\right]^2\rd
s\right)\rd y}.
\end{equation}
The unnormalised density---the numerator in
(\ref{altCondDens})---is for
all $x\in\R$ conditionally log-normal given $X_{}$. The simulation of the density requires only the
numerical implementation of the standard Brownian motion and of the random variable
$X$.
\end{rem}
\begin{rem}
The density models with deterministic volatility structure presented
in Proposition \ref{Prop inftime models} can be extended to a class of
models that satisfy the following system:
\\

Let $f_0(x):\R\rightarrow\R^+$ be a density function. A
filtered probability space
$\left(\Omega,\Fcal,\{\Fcal_t\},\Q\right)$ can be constructed along
with (i) an $\Fcal_{\infty}$-measurable random variable $X_{}$
with density $f_0(x)$, (ii) an $\{\Fcal_t\}$-adapted density process $\{f_t(x)\}$, and 
(iii) an $\{\Fcal_t\}$-adapted Brownian motion $\{W_t\}$,
such that (a) for some function $\gamma(t)$ on
$[0,\infty)$, (b) for some function $g(x)$ that is invertible onto
$\R$, and (c) for some suitably integrable function $v(t,x)$ on
$[0,\infty)\times\R$ with the properties
\begin{equation}
\lim_{t\rightarrow  \infty}\sqrt{t}\,\gamma(t)=0,\quad  \lim_{t \rightarrow \infty}\gamma(t)\int^t_0v(s,x)\rd s=g(x),
\end{equation}
the following relations hold for all $t\in[0,\infty)$. We have
$ \Q\left[X_{}\in \rd x\,\vert\,\Fcal_t\right]=f_t(x) \,\rd x $
and
\begin{equation}
f_t(x)=f_0(x)+\int^t_0
f_s(x)\left[v(s,x)-\int^{\infty}_{-\infty}v(s,y)f_s(y)\rd
y\right]\rd W_s.
\end{equation}
\end{rem}
\begin{rem}
The construction of conditional density models admits
an interpretation as a kind of a filtering
problem. The process (\ref{IDef}) plays the role of an ``observation process'', and
(\ref{XInnovations}) has the interpretation of being an ``innovation process''. See \cite{BE}, \cite {KR}, \cite{LS}.
\end{rem}
\begin{rem}
It is reasonable on a heuristic basis to expect that the
general deterministic volatility structure model can be calibrated
to the specification of an essentially arbitrary volatility surface.
In particular, the parametric freedom implicit in a deterministic
volatility structure coincides with that of a volatility surface.
The situation is rather similar to that of the relation arising in
the Dupire (1994) model between the local volatility (which is
determined by a deterministic function of two variables, one with the
dimensionality of price and the other with that of time) and the
initial volatility surface (which represents a two-parameter family
of option prices, labelled by strike and maturity). The precise
characterisation of such relations constitutes a non-trivial and
important inverse problem.
\end{rem}
\begin{rem}
In the general deterministic volatility structure model, the dynamics of the underlying asset price $\{A_t\}$ are of the form $\rd A_t= V_t\,\rd W_t$, with
\begin{equation}
V_t=\int_{\R} x\,v(t,x) f_t(x)\rd x-\int_{\R}x f_t(x)\rd x\int_{\R}v(t,x)f_t(x)\rd x, 
\end{equation}
where $f_t(x)$ is given by (\ref{PROPFCOND}). Thus the absolute volatility $V_t$ at time $t$ takes the form of a conditional covariance between $X$ and $v(t,X)$. In general, $\{V_t\}$ is an $\{\Fcal_t\}$-adapted stochastic volatility process that cannot be expressed in the form $\Sigma(t,A_t)$ for some function $\Sigma(t,x)$, and the dynamics do not constitute a simple diffusion of the Dupire (local volatility) type. In the ``semilinear"case $v(t,x)=\sigma T x/(T-t)$, however, the associated information process is Markovian, and $V_t$ can indeed be expressed in the form $\Sigma(t,A_t)$.
\end{rem}
\section{Semilinear volatility structure}\label{Sec 4}
We consider in this section the case where $A_T$ has a
prescribed unconditional density $f_{0T}(x)$, and construct a family of conditional density processes $\{f_{tT}\}$
that solve the master equation (\ref{ME}) over the time interval
$[0,T)$. The filtration with respect to which $\{f_{tT}\}$ is defined will be constructed as follows.
We introduce a process $\{\xi_{tT}\}_{0\le
t\le T}$ given by
$\xi_{tT}=\sigma\,A_T\,t+\beta_{tT}$,
where $\sigma$ is a constant and $\{\beta_{tT}\}_{0\le t\le T}$ is a
standard Brownian bridge, taken to be independent from $A_T$. We assume that $\{\Fcal_t\}$ is given by
\begin{equation}\label{xi-F-filtration}
\Fcal_t=\sigma\left(\left\{\xi_{sT}\right\}_{0\le s\le t}\right).
\end{equation}
Clearly $\{\xi_{tT}\}$ is $\{\Fcal_t\}$-adapted, and $A_T$ is
$\Fcal_T$-measurable. It is shown in Brody {\it et al.}~(2007, 2008) that $\{\xi_{tT}\}$ is an
$\{\Fcal_t\}$-Markov process (see also Rutkowski \& Yu 2007).
\begin{prop}\label{Prop BHM} {\rm Let the initial density $f_{0T}(x)$ be prescribed, and let the volatility structure
be of the semilinear form
\begin{equation}\label{Bachelier Vol Structure}
\sigma_{tT}(x)=\sigma\,\frac{T}{T-t}\,x,
\end{equation}
for $0\le t<T$, and let $\{\Fcal_t\}$ be defined by
(\ref{xi-F-filtration}). Then the process $\{W_t\}_{0\le t< T}$
defined by
\begin{equation}\label{SOL2}
W_t=\xi_{tT}-\int^t_0\frac{1}{T-s}\,\left(\sigma\,T\,\E\left[A_T\,\vert\,\xi_{sT}\right]-\xi_{sT}\right)\,\rd
s
\end{equation}
is an $\{\Fcal_t\}$-Brownian motion, and the process $\{f_{tT}(x)\}$,
given by
\begin{equation}\label{SOL1}
f_{tT}(x)=\frac{f_{0T}(x)\exp\left[\frac{T}{T-t}\left(\sigma\xi_{tT}x-\tfrac{1}{2}\,\sigma^2
x^2 t\right)\right]}
{\int^{\infty}_{-\infty}f_{0T}(y)\exp\left[\frac{T}{T-t}\left(\sigma\xi_{tT}y-\tfrac{1}{2}\,\sigma^2
y^2 t\right)\right]\rd y},
\end{equation}
satisfies the master equation (\ref{ME}) with the given initial
condition.}
\end{prop}
\noindent{\bf Proof}. The fact that $\{W_t\}_{0\le t< T}$ is an
$\{\Fcal_t\}$-Brownian motion is shown in Brody {\it et al.}~(2007,
2008). We work out $\E[A_T\,\vert\,\xi_{tT}]$ by use of the Bayes
formula,
\begin{equation}
\E\left[A_T\,\vert\,\xi_{tT}\right]=\int_{\R}x\,f_{tT}(x)\rd x,
\end{equation}
where
\begin{equation}
f_{tT}(x)=\frac{f_{0T}(x)\,\rho\left(\xi_{tT}\,\vert\,A_T=x\right)}{\int_{\R}f_{0T}(y)\,\rho\left(\xi_{tT}\,\vert\,A_T=y\right)\rd
y}.
\end{equation}
Here $\rho(\xi_{tT}\,\vert\,A_T=x)$ is the conditional density of
$\xi_{tT}$ given the value of $A_T$. We observe that $\xi_{tT}$ is conditionally Gaussian:
\begin{equation}
\rho\left(\xi_{tT}\,\vert\,A_T=x\right)=\sqrt{\frac{T}{2\pi\,
t(T-t)}}\,\exp\left[-\frac{1}{2}\frac{T}{t(T-t)}\left(\xi_{tT}-\sigma
t x\right)^2\right].
\end{equation}
Thus the density process is given by
\begin{equation}
f_{tT}(x)=\frac{f_{0T}(x)\exp\left[-\tfrac{1}{2}\,\frac{T}{t(T-t)}\left(\xi_{tT}-\sigma
tx\right)^2\right]}{\int^{\infty}_{-\infty}f_{0T}(y)\exp\left[-\tfrac{1}{2}\,\frac{T}{t(T-t)}\left(\xi_{tT}-\sigma
ty\right)^2\right]\rd y}.
\end{equation}
The last expression can be simplified after some rearrangement so as to take the form 
\begin{equation}
f_{tT}(x)=\frac{f_{0T}(x)\exp\left[\frac{T}{T-t}\left(\sigma\xi_{tT}x-\tfrac{1}{2}\,\sigma^2
x^2 t\right)\right]}
{\int^{\infty}_{-\infty}f_{0T}(y)\exp\left[\frac{T}{T-t}\left(\sigma\xi_{tT}y-\tfrac{1}{2}\,\sigma^2
y^2 t\right)\right]\rd y}.
\end{equation}
With this result at hand, we can write the process
$\{W_t\}_{0\le t<T}$ in the form
\begin{equation}
W_t=\xi_{tT}-\int^t_0\frac{1}{T-s}\left(\sigma
T\int_{\R}x\,f_{sT}(x)\rd x-\xi_{sT}\right)\rd s.
\end{equation}
We recall that the master equation (\ref{ME}) can be written as (\ref{f-intform}). We shall prove that (\ref{SOL1}) satisfies (\ref{ME}) by showing that (\ref{f-intform})
reduces to (\ref{SOL1}) if we insert (\ref{SOL2}) in
(\ref{Z-process}) and choose the volatility structure to be
(\ref{Bachelier Vol Structure}). For the process $\{Z_t\}$ in
(\ref{Z-process}) we obtain
\begin{equation}\label{ZW}
Z_t=\xi_{tT}+\int^t_0\frac{\xi_{sT}}{T-s}\,\rd s.
\end{equation}
The next step is to insert $\{Z_t\}$ in
the exponent
\begin{equation}\label{EXP}
\int^t_0\sigma_{sT}(x)\rd
Z_s-\tfrac{1}{2}\int^t_0{\sigma_{sT}}(x)^2\rd s
\end{equation}
appearing in equation (\ref{f-intform}). Expression (\ref{EXP}) can
be simplified by use of (\ref{Bachelier Vol Structure}) to give
\begin{align}\label{EXP result} \sigma\,T
x\int^t_0\frac{1}{T-s}\left(\rd\xi_{sT}+\frac{\xi_{sT}}{T-s}\,\rd
s\right)-\tfrac{1}{2}\,(\sigma\,T x)^2\int^t_0\frac{1}{(T-s)^2}\,\rd
s=\frac{T}{T-t}\left(\sigma x\xi_{tT}-\tfrac{1}{2}\,\sigma^2 x^2
t\right).
\end{align}
To derive this result, we make use of the relation
\begin{equation}
\int^t_0\frac{\rd\xi_{sT}}{T-s}=\frac{\xi_{tT}}{T-t}-\int^t_0\frac{\xi_{sT}}{(T-s)^2}\,\rd
s.
\end{equation}
With equation (\ref{EXP result}) at hand, we see that
(\ref{f-intform}) reduces to (\ref{SOL1}) if
(\ref{Bachelier Vol Structure}) holds. \hfill$\Box$
\section{Semilinear volatility: Brownian motion approach}
\noindent We proceed to show how the models constructed in Section
\ref{Sec 4} are related to the density models with
deterministic volatility structure treated in Section \ref{Sec Imp
Dens Mod}.
In particular, we consider a deterministic semilinear volatility function of the form
\begin{equation}\label{BHM vol}
v(t,x)=\sigma\,\frac{T}{T-t}\,x,
\end{equation}
where $0\le t<T$. For this volatility function the process
$\{I_t\}$ has the dynamics
\begin{equation}\label{EXII IPx}
\rd I_t=\sigma\,\frac{T}{T-t}\,X\,\rd t+\rd B_t.
\end{equation}
We are thus able to work out the exponent
\begin{equation}\label{EXII IP}
\int^t_0 v(s,x)\rd I_s-\tfrac{1}{2}\int^t_0 v^2(s,x)\rd s
\end{equation}
in equation (\ref{PROPFCOND}) making use of (\ref{BHM vol}) and
(\ref{EXII IPx}). We have:
\begin{align}\label{woEXP}
&\int^t_0 v(s,x)\rd I_s-\tfrac{1}{2}\int^t_0 v^2(s,x)\rd
s\nn\\
&=\frac{T}{T-t}\,\sigma\,x\left[(T-t)\int^t_0\frac{\rd
B_s}{T-s}+T(T-t)\,\sigma\,X\int^t_0\frac{\rd
s}{(T-s)^2}\right]-\tfrac{1}{2}T^2\sigma^2 x^2\int^t_0\frac{\rd
s}{(T-s)^2}.
\end{align}
The first integral gives rise to a $(\{\Gcal_t\},\Q)$-Brownian
bridge $\{\beta_{tT}\}$ over the interval $[0,T]$. More
specifically, we have
\begin{equation}
\beta_{tT}=(T-t)\int^t_0\frac{\rd B_s}{T-s}.
\end{equation}
The deterministic integral in (\ref{woEXP}) gives
\begin{equation}
\int^t_0\frac{\rd s}{(T-s)^2}=\frac{t}{T(T-t)}.
\end{equation}
Armed with these results, one can write (\ref{EXII IP}) as follows:
\begin{equation}\label{EXII EXP}
\int^t_0 v(s,x)\rd I_s-\tfrac{1}{2}\int^t_0 v^2(s,x)\rd
s=\frac{T}{T-t}\,\sigma\,x\left(\sigma\,X\,t+\beta_{tT}\right)-\frac{1}{2}\,\frac{T}{T-t}\,\sigma^2
x^2 t.
\end{equation}
Let $\{\xi_{tT}\}$ be defined for $t\in[0,T]$ by
$\xi_{tT}=\sigma\,X\,t+\beta_{tT}$.
Then for (\ref{EXII EXP}) we obtain
\begin{equation}\label{I&xi}
\int^t_0 v(s,x)\rd I_s-\tfrac{1}{2}\int^t_0 v^2(s,x)\rd
s=\frac{T}{T-t}\left(\sigma\,x\,\xi_{tT}-\tfrac{1}{2}\,\sigma^2 x^2
t\right).
\end{equation}
We conclude that the conditional density process $\{f_t(x)\}$ in
(\ref{PROPFCOND}) reduces to the following expression in the case
for which the volatility structure is given by
(\ref{BHM vol}):
\begin{equation}\label{BHM cond dens}
f_t(x)=\frac{f_0(x)\exp\left[\frac{T}{T-t}\left(\sigma\,x\,\xi_{tT}-\tfrac{1}{2}\,\sigma^2
x^2
t\right)\right]}{\int^{\infty}_{-\infty}f_0(y)\exp\left[\frac{T}{T-t}\left(\sigma\,y\,\xi_{tT}-\tfrac{1}{2}\,\sigma^2
y^2 t\right)\right]\rd y}.
\end{equation}
From equation (\ref{I&xi}) we see that $\{\xi_{tT}\}$
takes the role of the information process that generates
$\{\Fcal_t\}$. Since $\{\beta_{tT}\}$ vanishes for $t=T$, the random
variable $X$ is ``revealed'' at $T$. Thus $X$ is
$\Fcal_T$-measurable, and $\{\xi_{tT}\}$ is the process generating the
information-based models of Brody {\it et al.}~(2007, 2008). This
conclusion is supported by the following construction.

We consider the measure $\B$ defined in (\ref{B measure}). Under
$\B$ the process $\{I_t\}$ is a Brownian motion over the interval
$[0,T)$. We construct a $\B$-Brownian bridge by use of the
$\B$-Brownian motion $\{I_t\}$ as follows. On $[0,T)$ we set
\begin{equation}
\xi_{tT}=(T-t)\int^t_0\frac{1}{T-s}\,\rd I_s.
\end{equation}
Next we recall definition (\ref{IDef}) and insert this in the
expression above. The result is
\begin{equation}
\xi_{tT}=(T-t)\int^t_0\frac{\rd
B_s}{T-s}+(T-t)\int^t_0\frac{1}{T-s}\,v(s,X)\,\rd s.
\end{equation}
The first integral defines a $(\{\Gcal_t\},\Q)$-Brownian bridge over the interval $[0,T)$ which we denote $\{\beta_{tT}\}$.
For the volatility function we set
\begin{equation}\label{BHM volatility}
v(t,x)=\sigma\,\frac{T}{T-t}\,x.
\end{equation}
This leads to
\begin{equation}
\xi_{tT}=\sigma\,X\,(T-t)T\int^t_0\frac{1}{(T-s)^2}\,\rd
s+\beta_{tT},
\end{equation}
and thence to (6.8).
\section{Bachelier model}
The Bachelier model is obtained by setting $A_t=\gamma W_t$ where $\gamma$ is a
constant. We shall show that the class of models defined by
Proposition \ref{Prop BHM} contains the Bachelier model. We
consider a random variable $A_T$ associated with a fixed date $T$.
We assume that $A_T\sim N[0,1/(T\sigma^2)]$, where $N[m,v]$ is
the class of Gaussian random variables with mean $m$ and
variance $v$. In the notation of Section \ref{Sec 4}, we have
\begin{equation}
f_{0T}(x)=\frac{\sigma\sqrt{T}}{\sqrt{2\pi}}\,\exp\left(-\tfrac{1}{2}\,\sigma^2\,
T x^2\right).
\end{equation}
We recall the process $\{\xi_{tT}\}_{0\le t\le T}$ defined by
$\xi_{tT}=\sigma\,A_T\,t+\beta_{tT}$. If $A_T\sim N(0,1/(T\sigma^2))$,
then $\{\xi_{tT}\}$ is an $\{\Fcal_t\}$-Brownian motion over
$[0,T]$. This is because $\{\xi_{tT}\}$ is a
continuous Gaussian process with $\xi_{0T}=0$ and
$\textrm{Cov}[\xi_{sT},\xi_{tT}]=s$ for $0\le s\le t\le T$. We
recall the definition of the Brownian motion $\{W_t\}$ associated
with $\{\xi_{tT}\}$, given by (\ref{SOL2}). Since for $A_T\sim
N(0,1/T\sigma^2)$ the process $\{\xi_{tT}\}$ is a Brownian motion,
it follows that
\begin{equation}
\E\left[A_T\,\vert\,\xi_{sT}\right]=\frac{1}{\sigma
T}\,\E\left[\xi_{TT}\,\vert\,\xi_{sT}\right]=\frac{1}{\sigma
T}\,\xi_{sT}.
\end{equation}
Thus we see that $W_t=\xi_{tT}$. As a consequence we have
\begin{equation}
A_t=\E\left[A_T\,\vert\,\xi_{tT}\right]=\frac{1}{\sigma
T}\,\E\left[W_T\,\vert\,W_t\right]=\frac{1}{\sigma T}\,W_t.
\end{equation}
Hence to match the Bachelier model with an element in the class of
models constructed in Section \ref{Sec 4}, it suffices to set
$\sigma=1/(\gamma\,T)$.
\begin{prop}\label{Prop Bachelier}{\rm The conditional density process
$\{f^B_{tT}(x)\}$ of the Bachelier price process, defined over the
interval $[0,T)$, given by
\begin{equation}\label{Bachelier cond dens}
f^B_{tT}(x)=\frac{\exp\left[-\tfrac{1}{2}\frac{1}{\gamma^2(T-t)}(x-\gamma
W_t)^2\right]}
{\int^{\infty}_{-\infty}\exp\left[-\tfrac{1}{2}\frac{1}{\gamma^2(T-t)}(y-\gamma
W_t)^2\right]\rd y},
\end{equation}
is a special case of the family of the models of Proposition \ref{Prop
BHM}, and is obtained by setting
\begin{align}\label{Bachelier ini dens}
&f_{0T}(x)=\frac{\sigma\sqrt{T}}{\sqrt{2\pi}}\,\exp\left(-\tfrac{1}{2}\,\sigma^2\,
T x^2\right),& &\sigma_{tT}(x)=\sigma\,\frac{T}{T-t}\,x,& &
\textrm{and}& &\sigma=1/(\gamma\,T).&
\end{align}
}
\end{prop}
\noindent
{\bf Proof}. We insert (\ref{Bachelier ini dens}) in (\ref{SOL1}).
Completion of squares gives
\begin{align}
f_{tT}(x)&=\frac{f_{0T}(x)\exp\left[\frac{T}{T-t}\left(\sigma\xi_{tT}x-\tfrac{1}{2}\,\sigma^2
x^2 t\right)\right]}
{\int^{\infty}_{-\infty}f_{0T}(y)\exp\left[\frac{T}{T-t}\left(\sigma\xi_{tT}y-\tfrac{1}{2}\,\sigma^2
y^2 t\right)\right]\rd y}\nn\\
\nn\\
&=\frac{\exp\left[-\tfrac{1}{2}\frac{\sigma^2
T^2}{T-t}\left(x-\frac{1}{\sigma\,T}\,\xi_{tT}\right)^2\right]}{\int^{\infty}_{-\infty}\exp\left[-\tfrac{1}{2}\frac{\sigma^2
T^2}{T-t}\left(y-\frac{1}{\sigma\,T}\,\xi_{tT}\right)^2\right]\rd
y}.
\end{align}
Recalling that $\xi_{tT}=W_t$, and setting $\sigma=1/(\gamma\,T)$,
we obtain the desired result.\hfill$\Box$
\begin{rem}\label{Bachelier Corollary}{\rm Let the initial density
$f_{0T}(x)$, the volatility structure $\sigma_{tT}(x)$, and the
parameter $\sigma$ be given as in (\ref{Bachelier ini dens}). Then
the Bachelier conditional density $\{f^B_{tT}(x)\}$ satisfies the
master equation (\ref{ME}), and $\{W_t\}$
coincides with $\{\xi_{tT}\}$.}
\end{rem}
\begin{rem}
Suppose we chose an asset price model with a certain law. Then we know that we can derive the corresponding conditional density process where the related volatility structure and initial density are specified. We may then wonder how the conditional density transforms, and what the new volatility structure looks like, if we consider a new law for the asset price model. For instance, we may begin with the Bachelier model and ask what is the conditional density and volatility structure associated with a log-normal model.
We present a ``transformation formula'' for the conditional density. This result allows for the construction of a variety of conditional density processes from a given one.  Let $\{f_t(x)\}$ solve (\ref{ME}), and let $\psi:\R\rightarrow\R$ be a $C^1$-bijection. Then it is known that if $X$ has conditional density $f_t(x)$ then $Z=\psi(X)$ has conditional density $g_t(z)$ given by
\begin{equation}\label{transformformula}
g_t(z)=\frac{f_t\left(\psi^{-1}(z)\right)}{\psi'\left(\psi^{-1}(z)\right)},
\end{equation}
where $\psi^{-1}(z)$ is the inverse function and $\psi'(x)$ is the derivative of $\psi(x)$. The conditional density $\{g_t(z)\}$ satisfies
\begin{equation}
 \rd g_t(z)=g_t(z)\left[\nu(t,z)-\langle\nu_t\rangle\right]\rd W_t,
\end{equation}
where $\nu(t,z)=v(t,\psi^{-1}(z))$ and
\begin{equation}
 \langle\nu_t\rangle=\int_{\R}v(t,y)g_t(y)\rd y.
\end{equation}
We see that the volatility structure $v(t,x)$ associated with $f_t(x)$ becomes the volatility structure $\nu(t,z)$ associated with $g_t(z)$.
For example, consider the Bachelier model where $A_T\sim N(0,1/(T\sigma^2))$. The associated initial density and volatility structure are given in (\ref{Bachelier ini dens}). Now suppose that $Z=\exp(A_T)$, so $\psi(x)=\exp(x)$. The price process $\{A_t\}$ is then given by the log-normal model
\begin{equation}\label{log-normal model}
 A_t=\exp(\gamma W_t),
\end{equation}
where $W_t=\xi_{tT}$. It follows by (\ref{transformformula}) that the conditional density process $\{g_{tT}(x)\}$ associated with the log-normal price process (\ref{log-normal model}) is
\begin{equation}
 g_{tT}(z)=\frac{\exp\left[-\tfrac{1}{2}\frac{1}{\gamma^2(T-t)}\left(\ln(z)-\gamma W_t\right)^2\right]}{z\int^{\infty}_0\exp\left[-\tfrac{1}{2}\frac{1}{\gamma^2(T-t)}\left(\ln(y)-\gamma W_t\right)^2\right]\rd y},
\end{equation}
for $z>0$. Indeed we see that $g_{tT}(z)$ is the log-normal conditional density. The associated volatility structure is
\begin{equation}
 \nu(t,z)=v(t,\ln(z))=\sigma_{tT}(z)=\sigma\,\frac{T}{T-t}\,\ln(z).
\end{equation}
\end{rem}
\section{Option prices}
We consider a European-style call option with maturity $t$, strike
$K$, and price as determined by equation (\ref{svalueoption}). The price process $\{A_t\}_{0\le t<\infty}$ of the underlying asset
is given by
\begin{eqnarray}\label{underlier}
A_t&=&\E^{\Q}\left[X\,\vert\,\Fcal_t\right]
=\frac{\int^{\infty}_{-\infty}x\,f_0(x)\exp\left(\int^t_0
v(s,x)\rd I_s-\tfrac{1}{2}\int^t_0 v^2(s,x)\rd s\right)\rd
x}{\int^{\infty}_{-\infty}\,f_0(y)\exp\left(\int^t_0 v(s,y)\rd
I_s-\tfrac{1}{2}\int^t_0 v^2(s,y)\rd s\right)\rd y},
\end{eqnarray}
where  $\{\Fcal_t\}_{0\le s\le t<\infty}$ is generated
by (\ref{IDef}). We recall that $\{B_t\}$ is a $(\{\Gcal_t\},\Q)$-Brownian motion, and that there exists an
$(\{\Fcal_t\},\Q)$-Brownian motion $\{W_t\}$ such that
\begin{equation}\label{dIdW}
\rd I_t=\rd W_t\ +\langle v_t\rangle\rd t,
\end{equation}
where the bracket notation is defined by (\ref{bracketnotation}). We introduce a positive $(\{\Fcal_t\},\Q)$-martingale
\begin{equation}
\Lambda_t=\exp\left(\int_0^t\langle v_s\rangle\rd
W_s+\tfrac{1}{2}\int^t_0\langle v_s\rangle^2\rd s\right),
\end{equation}
which induces a change of measure from $\Q$ to measure
$\Q^{\ast}$ given by
\begin{equation}
\frac{\rd\Q^{\ast}}{\rd\Q}\bigg\vert_{\Fcal_t}=\Lambda_t \, .
\end{equation}
The $\Q^{\ast}$-measure is characterised by the fact that $\{I_t\}$
is an $(\{\Fcal_t\},\Q^{\ast})$-Brownian motion. We
observe that by the relationship
\begin{align}\label{duality}
&\int_{\R}f_{0}(x)\exp\left(\int^t_0 v(s,x)\rd
I_s-\tfrac{1}{2}\int^t_0v^2(s,x)\rd s\right)\rd x\nn\\
&\hspace{5cm}=\exp\left(\int^t_0\langle v_s\rangle\rd
I_s-\tfrac{1}{2}\int^t_0\langle v_s\rangle^2\rd s\right)\\
&\hspace{5cm}=\Lambda_t\label{IW} \, ,
\end{align}
we can use the denominator in (\ref{underlier}) to write the option price $C_{st}$ in terms of a
conditional expectation taken with respect to
$\Q^{\ast}$ under which $\{I_t\}$ is a Brownian motion.  Equation
(\ref{IW}) is obtained by applying the relationship (\ref{dIdW}). We
then have
\begin{eqnarray}
C_{st}&=&\E^{\Q}\left[\left(A_t-K\right)^+\,\vert\,\Fcal_s\right]\\
&=&\E^{\Q}\left[\left(N_t\Lambda_t^{-1}-K\right)^+\,\bigg\vert\,\Fcal_s\right]\\
&=&\E^{\Q}\left[\Lambda_t^{-1}\left(N_t-K\Lambda_t\right)^+\,\bigg\vert\,\Fcal_s\right]\\
&=&\Lambda_s^{-1}\,\E^{\Q^{\ast}}\left[\left(N_t-K\Lambda_t\right)^+\,\vert\,\Fcal_s\right],
\end{eqnarray}
where
\begin{eqnarray}
N_t&=&\int^{\infty}_{-\infty}x\,f_0(x)\exp\left(\int^t_0 v(s,x)\rd
I_s-\tfrac{1}{2}\int^t_0 v^2(s,x)\rd s\right)\rd
x,\\
\Lambda_t&=&\int^{\infty}_{-\infty}\,f_0(x)\exp\left(\int^t_0
v(s,x)\rd I_s-\tfrac{1}{2}\int^t_0 v^2(s,x)\rd s\right)\rd x.
\end{eqnarray}
Since $\{I_t\}$ is an $(\{\Fcal_t\},\Q^{\ast})$-Brownian motion, the
conditional expectation simplifies to the calculation of a Gaussian
integral provided the zero of the max function can be computed. 

With these formulae in hand, we observe that a closed-form expression for the conditional
expectation can be worked out in the case of a binary initial density function of the form
\begin{equation}
f_0(x)=q_1\,\delta(x-x_1)+q_2\,\delta(x-x_2).
\end{equation}
Here $\delta(x)$ is the Dirac distribution and
$q_i=\Q[X_{}=x_i]$ for $i=1,2$. It follows that
\begin{equation}
N_t=\sum^2_{i=1}x_i\,q_i\,\EM_t(x_i), \quad \quad
\Lambda_t=\sum^2_{i=1}q_i\,\EM_t(x_i), \label{LamE}
\end{equation}
where we introduce the process
\begin{equation}
\EM_t(x_i)=\exp\left[\int^t_0 v(s,x_i)\rd I_s-\tfrac{1}{2}\int^t_0
v^2(s,x_i)\rd s\right].
\end{equation}
In the case where the random variable $X_{}$ takes the 
values $x_1$ and $x_2$, the option price is
\begin{equation}\label{Qstar Call}
C_{st}=\Lambda^{-1}_s\,\E^{\Q^{\ast}}_s\left[\left(\sum^2_{i=1}\left(x_i-K\right)q_i\,\EM_t(x_i)\right)^+\,\bigg\vert\,\Fcal_s\right].
\end{equation}
We observe that $\{\EM(x_i)\}$ is positive and has the
property $\E^{\Q^{\ast}}[\EM_t(x_i)]=1$. In particular, we can use $\{\EM_t(x_1)\}$ to define a change of measure from
$\Q^{\ast}$ to a new measure $\tilde{\Q}$ by setting
\begin{equation}
\frac{\rd\tilde{\Q}}{\rd\Q^{\ast}}\bigg\vert_{\Fcal_t}=\EM_t(x_1),
\end{equation}
together with $\rd I_t=\rd\tilde{I}_t+v(t,x_1)\,\rd t$. Next we pull
$\EM_t(x_1)$ out to the front of the max function in equation
(\ref{Qstar Call}) to obtain
\begin{equation}
C_{st}=\Lambda^{-1}_s\,\E^{\Q^\ast}\left[\EM_t(x_1)\left(q_1(x_1-K)+q_2(x_2-K)\frac{\EM_t(x_2)}
{\EM_t(x_1)}\right)^{+}\,\bigg\vert\,\Fcal_s\right].
\end{equation}
By use of the Bayes formula we express the option price in
terms of $\tilde{\Q}$\,:
\begin{equation}
C_{st}=\Lambda^{-1}_s\,\EM_s(x_1)\E^{\tilde{\Q}}\left[\left(q_1(x_1-K)+q_2(x_2-K)\frac{\EM_t(x_2)}
{\EM_t(x_1)}\right)^{+}\,\bigg\vert\,\Fcal_s\right].
\end{equation}
For the sake of a simplified notation we define
$R_{0t}=\EM_t(x_2)/\EM_t(x_1)$, and we observe that the process
$\{R_{0t}\}$ is an exponential
$(\{\Fcal_t\},\tilde{\Q})$-martingale:
\begin{equation}
R_{0t}=\exp\left(\int^t_0[v(s,x_2)-v(s,x_1)]\rd\tilde{I}_s-\tfrac{1}{2}\int^t_0[v(s,x_2)-v(s,x_1)]^2\rd
s\right).
\end{equation}
We write $R_{0t}=R_{0s}R_{st}$ so that
\begin{equation}\label{Call RRR}
C_{st}=\Lambda^{-1}_s\,\EM_s(x_1)\E^{\tilde{\Q}}\left[\left(\,q_1(x_1-K)+q_2(x_2-K)R_{0s}R_{st}\,\right)^{+}\,\big\vert\,\Fcal_s\right],
\end{equation}
and we note that $R_{0s}$ is $\Fcal_s$-measurable. One is thus left
with the task of finding the range of values of $R_{st}$ for
which the max function vanishes. Then we calculate the
Gaussian integral arising from the conditional expectation.
Recalling that $\{\tilde{I}_t\}$ is an
$(\{\Fcal_t\},\tilde{\Q})$-Brownian motion, we note that the
logarithm of $R_{st}$ is Gaussian. Let $Y$ be a standard Gaussian variable. Then we can write
\begin{equation}
\ln\left(R_{st}\right)=\sqrt{\int^t_s\left[v(u,x_2)-v(u,x_1)\right]^2\rd
u}\ \,Y-\tfrac{1}{2}\int^t_s[v(u,x_2)-v(u,x_1)]^2\,\rd u.
\end{equation}
By solving for
the logarithm of $R_{st}$ in the argument of the max function in (\ref{Call RRR}), we deduce that the max function is zero for
all values $y^{\ast}$ that $Y$ may take for which
\begin{equation}
y^{\ast}\le\frac{\ln\left[\frac{q_1(K-x_1)}{q_2(K-x_2)R_{0s}}\right]+\tfrac{1}{2}\int^t_s[v(u,x_2)-v(u,x_1)]^2\rd
u}{\sqrt{\int^t_s\left[v(u,x_2)-v(u,x_1)\right]^2\rd u}}.
\end{equation}
It follows therefore that the option price can be written in the form
\begin{align}
C_{st}=\Lambda^{-1}_s\,\EM_s(x_1)&\left[q_1(x_1-K)\frac{1}{\sqrt{2\pi}}\int^{\infty}_{y^{\ast}}\exp\left(-\tfrac{1}{2}\,y^2\right)\rd
y\right.\nn\\
&\hspace{2cm}\left.+q_2(x_2-K)R_{0s}\frac{1}{\sqrt{2\pi}}\int^{\infty}_{y^{\ast}}\exp\left(-\tfrac{1}{2}\,\eta^2(y)\right)\rd
y\right]\, ,
\end{align}
where
\begin{equation}
\eta(y)=y-\sqrt{\int^t_s\left[v(u,x_2)-v(u,x_1)\right]^2\rd u}\, .
\end{equation}
The two Gaussian integrals can be written in terms of the normal
distribution function $N(x)$. To highlight the similarity
with the Black-Scholes option price formula, we define
\begin{equation}
d^{-}_{st}=-y^{\ast}, \quad \quad
d^{+}_{st}=d^{-}_{st}+\sqrt{\int^t_s\left[v(u,x_2)-v(u,x_1)\right]^2\rd
u} \,,
\end{equation}
so that one can write
\begin{equation}
C_{st}=\Lambda^{-1}_s\,\EM_s(x_1)\left[\,q_1(x_1-K)\,N(d^{-}_{st})+q_2(x_2-K)R_{0s}\,N(d^+_{st})\,\right].
\end{equation}
We can simplify this expression further by use of (\ref{LamE}). Finally, we conclude that the price of the call option is given by the
following compact formula:
\begin{equation}
C_{st}=(x_1-K)\,\frac{q_1}{q_1+q_2\,R_{0s}}\,N(d_{st}^-)+(x_2-K)\,\frac{q_2}{q_1\,R^{-1}_{0s}+q_2}\,N(d^+_{st}).
\end{equation}
\newpage
\vskip 15pt \noindent {\bf Acknowledgements}. The authors thank T.
Bj\"ork, Y. Kabanov, M. Monoiyos, M. Schweizer, J. Sekine, J. Zubelli, and participants at the Sixth World Congress
of the Bachelier Finance Society for useful discussions. Part of this research has been carried out as part of the project ``Dynamic Asset Pricing'', National Centre of Competence in Research ``Financial Valuation and Risk Management'' (NCCR FINRISK), a research instrument of the Swiss National Science Foundation. LPH and AM are grateful to the Ludwig-Maximilians-Universit\"at M\"unchen and
the Fields Institute, Toronto, for hospitality. LPH acknowledges support from Lloyds TSB Bank Plc, Kyoto University, Shell UK Ltd, and the Aspen Center for Physics. AM acknowledges support by ESF through AMaMeF grant 2863, and thanks the
Vienna Institute of Finance and EPFL for the stimulating work environment. Part of this
research was carried out while AM was a member of the Department of Mathematics, ETH Z\"urich.

\vskip 15pt \noindent {\bf References}.

\end{document}